\documentclass[12pt]{article}
\usepackage{graphicx}
\usepackage{epsf} 
\newcommand{\ba}{\begin{array}}
\newcommand{\ea}{\end{array}}
\def\be{\begin{equation}}
\def\ee{\end{equation}}
\def\bea{\begin{eqnarray}}
\def\eea{\end{eqnarray}}
\def\bs{\begin{subequations}}
\def\es{\end{subequations}}

\begin{document}

\begin{flushright}  
BROWN-HET-1388 \\
MCGILL-04-03 \\
UdeM-GPP-TH-04-118\\
\end{flushright}

\begin{center}
\bigskip
{\Large \bf Dynamical CP Violation in the Early Universe} \\
\bigskip
\bigskip
{\large K.R.S. Balaji $^{a,}$\footnote{balaji@hep.physics.mcgill.ca},
Tirthabir Biswas  $^{a,}$\footnote{tirtho@hep.physics.mcgill.ca},
Robert H. Brandenberger $^{b,c,}$\footnote{rhb@het.brown.edu} \\ and
David London $^{a,d,}$\footnote{london@lps.umontreal.ca}}
\end{center}

\begin{flushleft}
~~~~~~~~~~~$a$: {\it Physics Department, McGill University,}\\
~~~~~~~~~~~~~~~{\it 3600 University St., Montr\'eal QC, Canada H3A 2T8}\\
~~~~~~~~~~~$b$: {\it Department of Physics, Brown University,
Providence, RI 02912, USA}\\
~~~~~~~~~~~$c$: {\it Perimeter Institute for Theoretical Physics,}\\
~~~~~~~~~~~~~~~{\it Waterloo ON, Canada, N2J 2W9}\\
~~~~~~~~~~~$d$: {\it Laboratoire Ren\'e J.-A. L\'evesque, 
Universit\'e de Montr\'eal,}\\
~~~~~~~~~~~~~~~{\it C.P. 6128, succ. centre-ville, Montr\'eal, QC,
Canada H3C 3J7}
\end{flushleft}

\begin{center} 
\bigskip (\today)
\vskip0.5cm
{\Large Abstract\\}
\vskip3truemm
\parbox[t]{\textwidth} {Following earlier ideas of Dolgov, we show
that the asymmetrical dynamical evolution of fields in the early
Universe provides a new source for CP violation. This can lead to
baryogenesis without any additional CP-violating interactions. The
magnitude of this CP violation is time-dependent. In particular, it
vanishes (or is very small) in the late Universe after the fields have
relaxed (or are in their final approach) to their vacuum values. We
provide an explicit example in which our mechanism is realized.}
\end{center}

\thispagestyle{empty}
\newpage
\setcounter{page}{1}
\baselineskip=14pt

\section{Introducing Dynamical CP Violation}

The observed CP violation in our Universe has so far only been
measured in the $K$-meson and $B$-meson sectors (see e.g.\
Ref.~\cite{CPrev} for recent reviews on the status of CP violation),
and is generally believed to be due to CP-violating phases in the
quark mass matrix (the Kobayashi-Maskawa (KM) mechanism \cite{KM}). CP
violation is one of the key criteria required in order to generate the
observed baryon-antibaryon asymmetry of the Universe starting with
symmetric initial conditions. However, because of the smallness of the
quark masses, CP violation from the KM mechanism is highly suppressed
for processes relevant to baryogenesis \cite{Jarlskog}, and all
successful mechanisms of baryogenesis studied to date postulate new
CP-violating couplings arising in new physics beyond the Standard
Model (for recent reviews of baryogenesis see e.g. Ref.~\cite{BGrev}).
On the other hand, the three required criteria for baryogenesis
\cite{Sakharov}, namely the existence of baryon number violating
processes, CP violation, and out-of-equilibrium dynamics, all are
present in the Standard Model.  Thus, one may wonder if it might not
be possible to realize successful baryogenesis without introducing new
sources of baryon number violation and new couplings which explicitly
break CP (for attempts in this direction see
Refs.~\cite{Farrar,Turok}).

In this Letter we point out that within early Universe cosmology there
exists a natural source for CP violation. This can be used to obtain
the enhanced CP violation required to make it possible to generate a
large enough baryon asymmetry in the context of Standard Model
baryogenesis. The key observation, already made some time ago by
Dolgov \cite{Dolgov}, is that in a model which contains several
complex scalar fields, initial conditions in a given small region of
the early Universe will typically generate an asymmetry in the phases
of the fields. This asymmetry can be initially induced by thermal or
quantum excitations of the fields about the symmetric state. A stage
of inflation in the early Universe will lead to an exponential
increase in the wavelength of the local fluctuation regions, thus
rendering our present Hubble patch of the Universe asymmetric.

If the asymmetry in the phases of the fields can be connected with a
CP asymmetry, then it is possible to realize a scenario in which the
Lagrangian is CP symmetric (modulo the CP-violating phases in the
quark mass matrix of the Standard Model), but the phase asymmetry of
the fields in the early Universe leads to (possibly large) CP
violation during the period when the fields are relaxing to their
ground state values (which we assume are symmetric)\footnote{With
respect to the use of rolling scalar fields, our scenario has a
certain analogy with the Affleck-Dine (AD) mechanism \cite{Affleck}.
However, while the AD mechanism involves new scalar fields carrying
baryonic charge and generating a net baryon number, our scalar fields
do not involve new baryon-number-violating processes. Note that
rolling scalar fields are also used in inflationary baryogenesis
scenarios \cite{DL,Hall,Kolb,Chung,Nano}.  Once again, in these
scenarios the Lagrangian contains new CP- or baryon-number-violating
interactions.}. Thus, a specific feature of our mechanism is that the
magnitude of CP violation is time dependent. Large CP violation in the
early Universe in sectors other than the KM mass matrix could thus be
compatible with the absence of such effects today. The largeness of CP
violation in the early Universe could then enhance the effectiveness
of various baryogenesis mechanisms which have been proposed. For
example, if the scalar fields are rolling at times corresponding to
the electroweak phase transition, the effectiveness of electroweak
baryogenesis could be significantly enhanced, maybe rendering it
possible to obtain baryogenesis with Standard Model physics which is
effective enough to generate the observed baryon to entropy ratio (see
Ref.~\cite{Nir} for an interesting but unrelated mechanism which can
increase the efficiency of CP violation in the early
Universe)\footnote{Note that the scalar fields can be identified with
moduli fields in higher-dimensional compactifications (see
Ref.~\cite{Pope} for a review), in particular the breathing and the
squashing modes which can also effect symmetry-breaking transitions in
gauge theories as they are evolving in time \cite{Moduli}.}.

The idea that CP violation might be large in the early Universe due to
charge-asymmetric initial conditions is due to Dolgov \cite{Dolgov},
who introduced the term {\it stochastic breaking of charge symmetry}
for it. It can be viewed as the realization of {\it spontaneous CP
violation} \cite{Lee} during a period in the early Universe. In the
following, we develop this idea further. We introduce a toy model to
demonstrate the viability of the mechanism. We then discuss the
specific processes which in our toy model can generate a net CP
asymmetry among the particle excitations of the fields, and
demonstrate that it is possible to transfer this asymmetry to a net
lepton number (which in turn can be transferred to a net baryon number
by local thermal equilibration). Note that phases in multi-Higgs
systems have also recently been invoked \cite{Kusenko} as a way of
generating the CP asymmetry in the early Universe necessary to see
resonantly-amplified baryogenesis.

\section{CP Violation in the Scalar Sector} 

We begin by introducing a toy model containing two complex scalar
fields $\phi_+$ and $\phi_-$. The fields are taken here to be
electrically neutral (the case of electrically charged scalar fields
will be analyzed in a separate publication \cite{Balaji}). The scalar
potential of the model is taken to be
\be
V(\phi_+,\phi_-) = \sum_{i=+,-} m_i^2 \phi_i^\dagger \phi_i +
V_4(\phi_+, \phi_-)~,
\label{pot}
\ee
where it is assumed that $m_+ > 2 m_-$, and
\be
V_4(\phi_+,\phi_-) = g \phi_-^\dagger \phi_-^\dagger \phi_- \phi_+ +
h.c.
\label{pot2}
\ee
The Lagrangian is taken to be CP-conserving, so that $g$ is real.

If the field initial conditions are asymmetric, as is expected in
cosmology, we now show that the above scalar field interactions can
generate a net CP asymmetry among the local field excitations. We will
assume that the cosmological initial conditions provide a phase
asymmetry in the two scalar fields $\phi_+$ and $\phi_-$. To be
specific, we assume that the initial value of $\phi_-$ is real, but
the initial value of $\phi_+$ has a phase $\alpha$. As we will see,
this yields the required weak phase necessary in order to generate a
net CP-asymmetry.

Inflationary cosmology \cite{Guth} provides a ready mechanism to
produce the kind of field initial conditions we
require\footnote{Although we discuss the origin of the weak phase
difference and of the excitation of the scalar fields in the context
of inflationary cosmology, the basic mechanism does not depend on
having a period of inflation. In the context of any cosmological
scenario with a hot initial phase, one would expect all fields to be
excited, and relative phases to be generated. Inflation provides a
natural mechanism for producing homogeneous initial conditions within
our Hubble patch.}. Assume that the masses $m_+$ and $m_-$ are light
compared to the Hubble expansion rate $H$ during inflation. Inflation
will exponentially red-shift the wavelength of quantum field
fluctuations in the two scalar fields $\phi_+$ and $\phi_-$. Since the
motion of the fields is highly over-damped, the field fluctuation
amplitude ${\cal A}$, which for quantum vacuum fluctuations will be of
order $H$ \cite{Linde,FordVil} and therefore large in comparison with
the scales of Standard Model physics, will hardly decrease (if the
initial perturbations are due to other than quantum vacuum
fluctuations, the field amplitudes will be even larger). Thus, at the
end of inflation, the fields $\phi_+$ and $\phi_-$ will have obtained
components $\phi_+^{(0)}$ and $\phi_-^{(0)}$ which are homogeneous on
the scale of the present Hubble radius and whose amplitudes are large
at late times. In general, there will be a relative phase difference
$\alpha$ between the two fields. Once the Hubble rate $H$ drops below
the value of the mass $m_i$, the quasi-homogeneous component of the
field $\phi_i$ ($i=+,-$) will begin to relax towards its vacuum value
$\phi_i = 0$. The calculations we perform below make use of an
approximation which is only valid in the later stages of the
relaxation of the scalar fields, namely when the fields 
have decreased to values smaller than the masses ($|\phi_i| < m_-$
for $i=+,-$). However, CP violation is taking place throughout
the relaxation process.

It is clear that our cosmological initial conditions contain an
asymmetry in the weak phases of the fields. However, we must show that
this phase asymmetry can be transformed into a CP asymmetry in the
localized excitations of the field (the field quanta) which are
produced during the relaxation process. In the following we will also
show that an asymmetry in the fermion sector is generated. In this
way, the asymmetry survives even after the two scalar fields have
relaxed to their ground state in which (by assumption) there is no
residual weak CP-violating phase.

Before doing this, it is useful to first review the basic formalism
for establishing a CP asymmetry. In general, one can parameterize the
transition matrix element $M$ corresponding to a particular decay as
\be
M = A + \zeta B ~,
\label{CPamp}
\ee
where $\zeta$, $A$ and $B$ are complex numbers such that under a CP
transformation we have $A\leftrightarrow A$, $B\leftrightarrow B$ and
$\zeta \to \zeta^*$. The probability difference between the process
and its CP-conjugate process can then easily be worked out, and the
result determines the resulting CP asymmetry $A_{CP}$:
\be
A_{CP} = C (|M|^2 -|\bar M|^2) = -2C {\rm Im}(\zeta)\cdot {\rm Im}(AB^*)~,
\label{acp}
\ee
with $C$ being a constant. 

Clearly, in order to obtain net CP violation, one needs both a
non-vanishing {\it strong phase} difference ${\rm Im}(AB^*) \neq 0$
and a {\it weak phase} ${\rm Im}(\zeta) \neq 0$. In our scenario, the
weak phase is given by the relative phase in the initial displacement
of the fields from their symmetric values, whereas the strong phase is
obtained in the usual way by making use of a radiative process such as
a self-energy correction or a vertex correction graph.

With this formalism in hand, we can now show that the asymmetric
initial conditions do indeed lead to a CP asymmetry in the fields.
Because of these nonzero initial conditions, we must consider field
fluctuations about these values:
\be
\phi'_- = \phi_- - c_- ~~,~~~~ \phi'_+ = \phi_+ - c_+ e^{i\alpha} ~,
\ee
where $c_-$ and $c_+$ are real constants. (Note that $c_{\pm}$
actually vary with time, but our analysis is performed at a given
instant. Apriori, even $\alpha$ can be time dependent. For our
mechanism to work we require it to be varying slowly (if at all) so
that the CP-violation effects can add coherently.) These relations can
be inverted and inserted into the scalar potential of Eq.~(\ref{pot}).
Expanding the quartic term of Eq.~(\ref{pot2}), one finds both
quadratic and cubic terms. (There are other terms, but they are not
relevant to our analysis.) The quadratic terms contribute to the mass
matrix. However, if we make the simplifying assumption (valid in the
later stages of the relaxation of the scalar fields) that $c_{+,-} \ll
m_{+,-}$, then the mass corrections due to the initial conditions are
negligible, and $\phi'_i$ are still mass eigenstates. (Note that this
assumption is not absolutely necessary: it is possible to diagonalize
the mass matrix and work with the new mass eigenstates, but this does
not change our conclusions.) We will henceforth drop the primes.

The cubic terms generated by the initial conditions, which we denote
as $V_3$, have the following form:
\be
V_3 = g \left[ c_+ e^{i\alpha} \phi_-^\dagger \phi_-^\dagger \phi_- +
c_- \phi_-^\dagger \phi_-^\dagger \phi_+ + 2 c_- \phi_-^\dagger \phi_-
\phi_+ \right] + h.c.
\label{cubic2}
\ee
These couplings can contribute to two-body decays of the fields, which
can be treated analogously to how the decay of the inflaton field was
initially analyzed perturbatively \cite{DL,AFW} \footnote{However, see
Refs.~\cite{TB90,KLS} for a more efficient decay mechanism making use
of a parametric resonance instability which occurs if the homogeneous
field is oscillating.}. Since $m_+ > 2 m_-$, the only possible decay
is of a $\phi_+$ quantum to two $\phi_-$ quanta. We need to show that
the decay produces a net CP asymmetry among the produced quanta.

\begin{figure}
\centerline{\epsfxsize=1.9 in \epsfbox{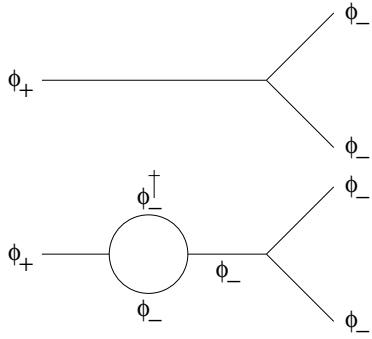}}
\caption{Feynman diagrams of the tree and loop processes considered here.}
\end{figure}

Consider the decay $\phi_+ \to \phi_- \phi_-$. This process receives a
tree and several loop contributions, with relative weak and strong
phases. For the purposes of demonstration, we will consider only one
of the diagrams which contributes at one loop (Fig.~1).  (There are
other loops of this type, as well as a vertex renormalization
graph.) The interference between these diagrams yields the desired CP
violation, and leads to a rate difference between $\phi_+ \to \phi_-
\phi_-$ and $\phi_+^\dagger \to \phi_-^\dagger \phi_-^\dagger$.

One can see how this comes about as follows. Using the couplings in
(\ref{cubic2}), we see that the tree and loop diagrams are
proportional to $c_- g$ and $2 c_- c_+^2 g^3 exp(2i\alpha)$,
respectively, and therefore have a relative weak phase. The loop
diagram also has a strong phase. Because the decay $\phi_+ \to \phi_-
\phi_-^\dagger$ is kinematically permitted, the loop transition
amplitude contains an absorptive part coming from the $i\epsilon$
piece in the propagators. Thus, we immediately see that the transition
matrix element for Fig.~1 will be of the necessary form
[Eq.~(\ref{CPamp})] to obtain net CP violation in the final
state. Specifically,
\be
A_{CP} \sim {\rm Im}(\zeta) \sim \sin 2\alpha ~.
\label{acps1}
\ee
We therefore see explicitly that $A_{CP}$ is nonzero only if
$\alpha\ne 0$. Thus, the phase $\alpha$ in the initial conditions is
directly responsible for the net CP asymmetry.

Note that we should also pay attention to the decay $\phi_+^\dagger
\to \phi_- \phi_-$. Even if there is a CP asymmetry between $\phi_+
\to \phi_- \phi_-$ and $\phi_+^\dagger \to \phi_-^\dagger
\phi_-^\dagger$, it could potentially be cancelled by another CP
asymmetry between $\phi_+^\dagger \to \phi_- \phi_-$ and $\phi_+ \to
\phi_-^\dagger \phi_-^\dagger$. However, since all triple-scalar
couplings can be different, it is not hard to ensure that this does
not arise. Indeed, our choice of $V_3$ achieves this: since there is
no $\phi_-^\dagger \phi_-^\dagger \phi_+^\dagger$ term, the decay
$\phi_+^\dagger \to \phi_- \phi_-$ cannot occur at tree level.

\section{Induced Lepton and Baryon Asymmetry} 

Above, we demonstrated that the initial phase asymmetry can be
converted to a CP-violating rate asymmetry in the scalar sector. The
next necessary step is to show how to transfer this to the fermion
sector. Given that the complex scalar fields $\phi_+$ and $\phi_-$ are
neutral, only a lepton asymmetry can be generated (through their
couplings to neutrinos). This then induces a net baryon asymmetry via
thermal equilibration. Thus, our mechanism can be a source for the
standard leptogenesis scenario (see Ref.~\cite{Fukugita} for the
original reference and e.g.\ Ref.~\cite{lepto} for a recent review of
leptogenesis).  However, special to our scenario is the fact that
leptogenesis can be driven by either by Dirac or by Majorana fermions
(not just by Majorana fermions as in the standard leptogenesis
scenario). (Note: if the scalar fields are charged, then their decay
can directly produce a baryon asymmetry (in this case the analogy with
the Affleck-Dine mechanism \cite{Affleck} would be greater).)

We first discuss the case where the fermions are Dirac particles. The
relevant Yukawa interaction is
\be 
Y = \sum_{a,b} ( \bar \psi^a_L (\phi_+ Y^{1}_{ab} + \phi_- Y^{2}_{ab})
\psi^b_R + h.c. \, ,
\label{yukawa}
\ee 
where the $Y^{i}_{ab}$ are the Yukawa coupling matrices, and the
indices $a$ and $b$ run over the neutrinos. The decay processes of
interest are $\phi_- \to \bar\psi_L \psi_R$ and $\phi_-^\dagger \to
\bar\psi_R \psi_L$. Since we have unequal numbers of $\phi_-$ and
$\phi_-^\dagger$ quanta [Eq.~(\ref{acps1})], the decays of the scalars
will lead to an unequal number of left- and right-handed Dirac
neutrinos. This asymmetry has magnitude $A^\nu \sim |Y|^2 A_{CP}$, and
can successfully generate a baryon asymmetry for very small Yukawa
couplings \cite{manfred}. There are, however, a number of qualitative
and quantitative differences between our scenario and that of
\cite{manfred}. These are described in detail in Ref.~\cite{Balaji}.

For the case of Majorana fermions $N_R$, one can simply use the usual
leptogenesis mechanism, with the scalars $\phi_\pm$ of our scenario
playing the role of the usual scalars of the standard leptogenesis
model \cite{Fukugita}. The difference between the two models is that,
in our scenario, the Yukawa couplings are real and CP violation is due
to the phase difference in the initial conditions of the scalar
fields.

There are several other ways of generating a lepton asymmetry in the
context of our proposed mechanism. Also, as mentioned earlier, it is
straightforward to have conventional baryogenesis via charged scalars. 
All of these scenarios will be discussed in Ref.~\cite{Balaji}.

\section{Discussion and Conclusions} 

In this Letter we have studied a mechanism for CP violation
(originally proposed by Dolgov \cite{Dolgov}) which is effective in
the early Universe but could shut off at late times. A key feature of
our mechanism is that it does {\it not} involve any new CP-violating
phases in the Lagrangian. Instead, in a model containing several
scalar fields, initial conditions will lead to an asymmetry in the
phases of these fields. CP-violating processes then arise as these
quantum fields relax dynamically towards their symmetric ground state
values. The initial displacement of these fields is completely natural
in the context of early Universe cosmology.

There are many possible applications for this mechanism. In this
paper, we have concentrated on its contribution to leptogenesis and
baryogenesis. Our calculation is applicable in the phase when the
fields are rolling. This rolling phase will start when the Hubble
constant drops to a value comparable to the mass of the scalar
fields. It is at this time in the cosmological evolution that CP
violation is most efficient. After the fields have relaxed to their
vacuum values, our CP violation mechanism turns off. We plan to
discuss more details, in particular applications to concrete
baryogenesis models, in a future publication \cite{Balaji}.  Note that
string cosmology and brane world scenarios may provide natural
settings for the origin of the scalar fields required for our
mechanism (e.g.\ see Ref.~\cite{Gian} for a recent paper on how scalar
fields from brane world scenarios can play a new role in spontaneous
baryogenesis).

\bigskip
\noindent
{\bf Acknowledgements}:
We thank Guy Moore and Cliff Burgess for useful discussions. One of us
(RB) is also grateful to Marco Peloso for interesting conversations.
RB is supported in part by the US Department of Energy under Contract
DE-FG02-91ER40688, TASK~A. He thanks the Fields Institute for
Mathematical Sciences and the Perimeter Institute for hospitality and
support during the final stages of this project. The work of KB, TB
and DL is supported by NSERC (Canada) and by the Fonds de Recherche
sur la Nature et les Technologies du Qu\'ebec.


\end{document}